\input{aipcheck}

\documentclass[
    ,final            
  ]
  {aipproc}

\layoutstyle{6x9}

\begin{document}

\title{First results from the NPDGamma experiment at the Spallation Neutron Source }

\classification{24.80.+y}
\keywords      {parity violation, fundamental symmetries, electroweak interactions}

\author{Nadia Fomin for the NPDGamma Collaboration}{
  address={Los Alamos National Laboratory, Los Alamos, NM, 87545, USA}
}

\begin{abstract}
 The NPDGamma experiment aims to measure the parity-odd correlation between the neutron spin and the direction of the emitted photon in neutron-proton capture.  A parity violating asymmetry (to be measured to 10$^{-8}$) from this process can be directly related to the strength of the hadronic weak interaction between nucleons, specifically the $\Delta$I=1 contribution.  As part of the commissioning runs on the Fundamental Neutron Physics beamline at the Spallation Neutron Source (SNS) at ORNL, the gamma-ray asymmetries from the parity-violating capture of cold neutrons on $^{35}$Cl and $^{27}$Al were measured, to check for systematic effects, false asymmetries, and backgrounds.  Early this year, the parahydrogen target for the production run of NPDGamma was commissioned.  Preliminary results for the commissioning measurements with  $^{35}$Cl and $^{27}$Al will be presented as well as first results of the hydrogen run.

\end{abstract}

\maketitle


\section{Introduction}
The $\Delta$S=1 piece of the hadronic weak interaction cannot be measured directly in the presence of the strong force.  Instead, we use the fact that the weak interaction violates parity, manifested in measurements of physics observables.  In the case of NPDGamma, we isolate the $\Delta I=$1 part of the interaction, resulting from the hadronic weak neutral current.  The measured asymmetry, $A_{\gamma}$ arises from the small admixture of $P$-wave states in the initial $S$-wave singlet and final $S$-wave triplet states.  The DDH model~\cite{ddh}, one of the the earliest ones for the hadronic weak interaction, 
uses valence quarks to calculate effective PV meson-nucleon coupling directly from SM via 7 weak meson coupling constants.  Physics observables are then written as their combinations, with the NPDGamma asymmetry given by:
\begin{equation}
A_{\gamma} = -0.1069h_{\pi}^1-0.0014h_{\rho}^1+0.0044h^1_{\omega},
\end{equation}
where the observed result is dominated by the long-range pion-nucleon interaction, meaning the asymmetry is predominantly sensitive to $h_{\pi}^1$.  The central value for the DDH model prediction is at $h_{\pi}^1$ = 4.7$\times$10$^{-7}$.  Recently, the first lattice QCD calculation of $h_{\pi}^1$ was carried out~\cite{Wasem:2011tp} using a pion mass of 589 MeV/c$^2$ and yielding a result of (1.099$\pm$0.505$^{+0.058}_{-0.064}$) $\times$10$^{-7}$ for the connected diagrams only.  NPDGamma will make a measurement of $h_{\pi}^1$ with a statistical precision of 1$\times$10$^{-8}$.
\section{Experimental Setup}
\begin{figure}[h!]
  \includegraphics[height=.25\textheight]{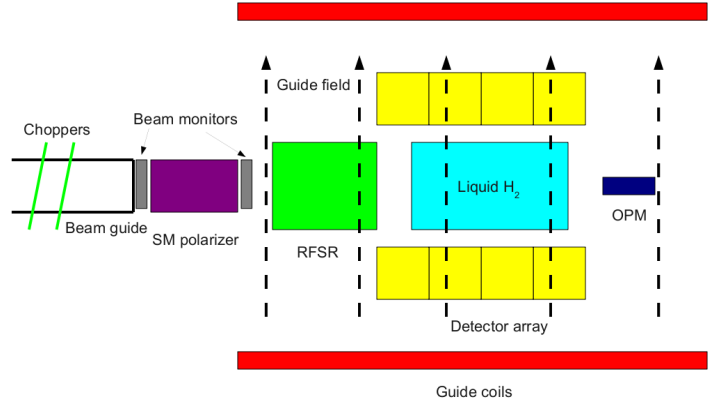}
  \caption{Left-Right schematic view of the NPDGamma experiment.  Cold neutrons pass through a supermirror polarizer (SMP), their spin is possibly reversed via a RFSR based on an alternating 8-step sequence, and they're incident on a 16 L liquid parahydrogen target.  Gamma rays from the capture are detected by an $\approx$ 3$\pi$ acceptance detector array of CsI(Tl) crystals.   The detectors are operated in current mode due to extremely high rates. A 9.4 G magnetic field maintains the spin of the neutrons.}
\label{fig1}
\end{figure}

The NPDGamma experiment is located $\approx$17.6 m (center of the target) downstream of the moderator at the SNS. A pair of bandwidth choppers is used to select neutrons with energies below 14.7 meV.  The pulsed cold neutron beam is spin-filtered by a supermirror polarizer (SMP)~\cite{Mezei:1976,Turchin}, yielding a highly polarized neutron beam ($<$95\%).  A resonant radio frequency spin rotator (RFSR)~\cite{Seo:2007yk} is used to reverse the spins of the neutrons according to an 8-step spin sequence ($\uparrow\downarrow\downarrow\uparrow\downarrow\uparrow\uparrow\downarrow$) or its inverse, chosen to minimize effects from time-dependent gain drifts to second order.  Additionally, this allows to to cancel differences from detector efficiencies in the analysis procedure.  Polarized neutrons are captured on protons in the 16 L target of liquid parahydrogen.  Neutrons with energies below 14.7 meV do not depolarize through scattering from the parahydrogen, as they would from orthohydrogen.  The resulting gamma rays are detected in the CsI detectors arranged around the target.  The experimental setup is shown in Fig.~\ref{fig1}.

\section{Summary of Preliminary Results}
In the analysis, asymmetries are formed for each spin sequence (octet), using integrated yields in the CsI detectors corresponding to captures of neutrons with opposite spin states.  As detectors are arranged in $\approx$3$\pi$ around the target, each detector will have a given sensitivity to the PV up-down and PC left-right asymmetries, given by $G_{UD}$ and $G_{LR}$.  In the case of negligible dilution, these so-called ``geometric factors'' are used to extract the physics asymmetries via
\begin{equation}
A_{raw}=A_{ud}\times G_{UD}+A_{lr}\times G_{LR}
\end{equation}

\begin{figure}[h!]
  \includegraphics[height=.35\textheight, angle=270]{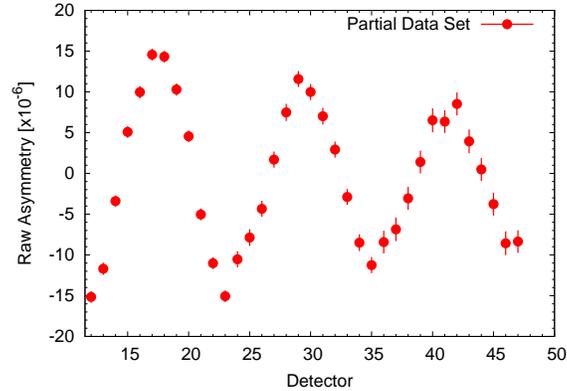}
  \caption{Raw Chlorine asymmetry extracted for each detector used.  The shape as a function of detector reflects the changing sensitivity to the up-down PV asymmetry.}
\label{fig2}
\end{figure}

Data have been collected on several targets.  A chlorine target was used because of its large and well-known asymmetry~\cite{Vesna:1982pp,Mitchell:2004fn,Avenier:1985xu}, with a current world average of -23.9$\pm$1.36 [x10$^{-6}$].  In addition to making a new and precise measurement, this target allows us to test the apparatus and analysis tools.  The raw chlorine asymmetry is shown in Fig.~\ref{fig2}, with the shape characteristic of the detectors' sensitivities.

\begin{figure}[h!]
\includegraphics[height=.3\textheight]{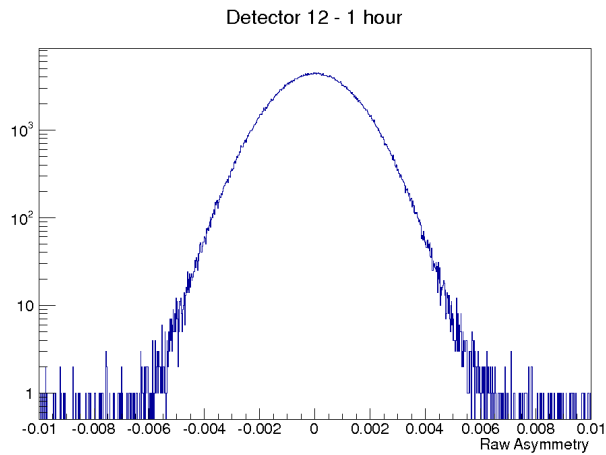}
\caption{Histogram of raw hydrogen asymmetries for a typical detector for 1 hour of data taking.}
\label{fig3}
\end{figure}

Additionally, data were also taken on an aluminum target, which is the largest source of background for NPDGamma.  In addition to a parity violating asymmetry from neutron capture on aluminum, it is a source of prompt and $\beta$-delayed background from the decay of $^{28}$Al$^{*}$ to $^{28}$Al.  Finally, production running with the hydrogen target is underway, with $\approx$5$\times$10$^{-8}$ statistics collected during the spring of 2012.  An example histogram of the raw hydrogen asymmetry for a typical detector is shown in Fig.~\ref{fig3}, for just one hour of data taking. Additive and multiplicative false asymmetries have been measured and found consistent with zero at 2$\times$10$^{-9}$ for this first run, and the backgrounds calculated using MCNPX have been confirmed experimentally.  The final phase of the experiment is currently under way, with data taking on hydrogen that will take us to our final sensitivity of 1$\times$10$^{-8}$.

\bibliographystyle{aipproc}   

\bibliography{325_fomin}

\IfFileExists{\jobname.bbl}{}
 {\typeout{}
  \typeout{******************************************}
  \typeout{** Please run "bibtex \jobname" to optain}
  \typeout{** the bibliography and then re-run LaTeX}
  \typeout{** twice to fix the references!}
  \typeout{******************************************}
  \typeout{}
 }

\end{document}